\documentclass[showpacs,preprintnumbers,amsmath,amssymb]{revtex4}
\usepackage{graphicx}
\usepackage{bm}
\usepackage{amsfonts}
\usepackage{indentfirst}
\usepackage[small]{caption2}
\usepackage{longtable}

\begin{document}


\title{Entanglement induced by a two-mode thermal field }
\author{E.K. Bashkirov}
 \altaffiliation[Electronic address:]{bash@ssu.samara.ru}
\affiliation{%
Department of General and Theoretical Physics, Samara State University, Acad. Pavlov Str.1 , 443011 Samara, Russia
\\
}%

\date{\today}

 \begin{abstract}
The entanglement between two identical two-level atoms interacting with two mode thermal field through a nondegenerate two-photon process
 has been suggested. It has been shown that for some atomic initial state the entanglement induced by nondegenerate
 two-photon interaction is larger than that induced by one-photon and degenerate two-photon processes.
\end{abstract}


\maketitle

The quantum entanglement plays a key role in quantum information and quantum computation \cite{Nielsen}.
 Real quantum systems will inevitably be influenced by surrounding environments. Although the interaction
 between the environment and quantum systems can lead to decoherence, it can also induced entanglement. Recently, S.Bose et al.
 \cite{Bose} have showed that entanglement can always arise in the interaction of an arbitrary large system in any
 mixed state with a single qubit in a pure
 state,  and illustrated this using the Jaynes-Cummings interaction of a two-level atom in a pure state with a field in a thermal state at an
 arbitrary high temperature. M.Kim et al. \cite{Kim} have investigated the entanglement of two identical
  two-level atoms with one-photon  transition induced by a single-mode
 thermal field. L.Zhou et al \cite{Zhou1} have considered the same problem for nonidentical atoms. The entanglement of
 two identical
 two-level atoms through  nonlinear two-photon interaction with one-mode thermal field has been studied by L.Zhou
 et al. \cite{Zhou2}. They have shown that atom-atom entanglement induced by nonlinear interaction is larger than that induced by linear interaction.
 It is of interest to investigate the entanglement between two identical atoms induced by
 thermal noise through other nonlinear processes. In this paper we dwell on two-photon nondegenerate
 interaction of two atoms with two mode of thermal field. The one-atom models with nondegenerate two-photon interaction
 in cavity have attracted
 much attention beginning with the paper of S.Gou \cite{Gou}. We have obtained that for some atomic initial
 condition the entanglement
induced by nondegenerate
 interaction may be  larger than that induced by  degenerate two-photon process.

Let us consider the system of two identical two-level atoms with frequency of the atomic transition $\omega_0$ interacting with two modes
 of quantum electromagnetic field
 with frequencies  $\omega_1$ and $\omega_2$ through a nondegenerate two-photon process. For simplicity we ignore the Stark
 shift and assume the lossless
 cavity modes  to be at two-photon resonance with the atomic transition, i.e.
 the condition $\omega_1 + \omega_2 = \omega_0$ takes place. Then, the Hamiltonian
 of the considered system in interaction picture and RWA approximation can be written in the form:
\begin{eqnarray}
H_{I} = \hbar g \sum\limits_{i=1}^2 (a^+_1 a^+_2R^-_i + R^+_i a_1 a_2),
\end{eqnarray}
where $a^+_j$ and $a_j$ are the creation and annihilation operators of photons of $j$th cavity mode ($j=1,2$),
 $R^+_i$ and $R^-_i$ are the raising and the lowering operators for the $i$th atom,
 $g$ is the coupling constant between the  atom and the cavity field.

We denote by  $\mid + \rangle$ and $\mid - \rangle$ the excited and ground states of a single two-level atom and by
 $\mid n\rangle$ the Fock state of the electromagnetic field. The two-atom wave function can be expressed as
 a combination of state vectors of the form $\mid \it v_1, \it v_2 \rangle = \mid \it v_1\rangle \mid \it v_2 \rangle $,
 where $\it v_1, \it v_2 = +,-$.

 The density operator for the atom-field system follows a unitary time evolution generated by the evolution operator
 $U(t) = \exp(-\imath H t/\hbar)$. In the two-atom basis $\> \mid +, + \rangle, \,\mid +, - \rangle, \,\mid -, + \rangle, \,
 \mid -, - \rangle$ the analytical form of the evolution operator $U(t)$ is given by
\begin{eqnarray} U(t) =\begin{pmatrix}
 a_1 a_2 C a^+_1 a^+_2 + 1 & -\imath a_1 a_2 S & -\imath a_1 a_2 S & a_1 a_2 C a_1 a_2\\
 -\imath a^+_1 a^+_2 S & \frac{1}{2}(\cos \Omega t +1) &\frac{1}{2}(\cos \Omega t -1)&-\imath S a_1 a_2  \\
 -\imath a^+_1 a^+_2 S & \frac{1}{2}(\cos \Omega t -1) &\frac{1}{2}(\cos \Omega t +1)&-\imath S a_1 a_2  \\
 a^+_1 a^+_2 C a^+_1 a^+_2  & -\imath a^+_1 a^+_2 S & -\imath a^+_1 a^+_2 S & a^+_1 a^+_2 C a_1 a_2 +1
\end{pmatrix}\,,\quad
\end{eqnarray}

where $$\Omega = g\sqrt{2(a_1 a_2 a^+_1 a^+_2 + a^+_1 a^+_2 a_1 a_2)}$$
 and
$$ C=\frac{2 g^2}{\Omega^2}(\cos \Omega t -1), \quad S =  \frac{g}{\Omega}\sin\Omega t. $$

 Let the cavity field is initially in two-mode thermal state with mean photon numbers $\overline {n}_1$ and $\overline {n}_2$. The thermal radiation field is a weighted mixture of Fock states
 for each modes and its density operator is represented by
$$\rho_F(0) = \sum\limits_{n_1\,n_2} p_{n_1}p_{n_2} \mid n_1 \rangle \langle n_1\mid \mid n_2 \rangle \langle n_2\mid,  $$
where the weight functions $p_{n_i}$ is
$$p_{n_i} = \frac{\overline {n}^{n_i}_i}{(1+\overline {n_i})^{n_i+1}} \qquad (i=1,2) $$
and the mean photon number $\overline {n}_i$ in the $i$th cavity mode ($i=1,2$) is $\overline {n}_i = (\exp(\hbar \omega_i/k_B T) - 1)^{-1}$, where $T$ is the equilibrium cavity temperature and $k_B$ is the Boltzmann constant.

To investigate the entanglement between atoms one can obtain the time-dependent reduced atomic density operator
 by means of the combined atom-field density operator over the field variables:
 \begin{eqnarray}
 \rho_A(t) = Tr_F U(t) \rho(0) U^+(t).
 \end{eqnarray}

 If the atoms are initially in a pure state such as $\> \mid +, + \rangle, \,\mid +, - \rangle, \,\mid -, + \rangle, \,$ or $\,
 \mid -, - \rangle$, the atomic density operator with using formulae (2), (3) may be expressed in a
 similar way to the nondegenerate two-photon case as has been obtained earlier by L.Zhou et al. \cite{Zhou2}

\begin{eqnarray}
\rho_A(t)=\begin{pmatrix}
 A & 0 & 0 &0\\
  0 & B & E &0\\
   0 & E & C &0\\
    0 & 0 & 0 &D
\end{pmatrix}.
\end{eqnarray}
For two-qubit system described by density operator $\rho_A(t)$, a measure of entanglement can be defined in
 terms of the negative eigenvalues
of partial transposition \cite{Peres}, \cite{Horodecki}
$$\varepsilon = -2 \sum\limits_i \mu_i^-, $$
where $\mu_i^-$ are the negative eigenvalues of the partial transposition of $\rho_A(t)$.
 When $\varepsilon = 0$ the two atoms are separable  and $\varepsilon >0 $ means the atom-atom
 entanglement. The case $\varepsilon = 1$ indicates maximum entanglement. For partial transposition of (4)
  we have only one eigenvalue which can be negative. This value is
$\mu_1^- = \frac{1}{2} (D +A - \sqrt{(D-A)^2 + 4 E^2)}$. The
eigenvalue $\mu^-_1$ becomes negative if and only if $ E >
\sqrt{A D}$. Under this condition, the measure of entanglement is

\begin{eqnarray}
\varepsilon = \sqrt{(D-A)^2 + 4 E^2)} -D - A.
\end{eqnarray}

Let us consider the time-dependent measure of entanglement for various initial pure states of the atomic subsystem.

 1. Let atoms are initially in the state $\mid +, - \rangle $. Then,  the matrix elements of the atomic reduced density  operator (4) are
\begin{eqnarray}
\nonumber A= \sum\limits_{n_1\,n_2} p_{n_1} p_{n_2} n_1 n_2 S^2_{n_1,\,n_2}\>,\\
\nonumber B=\sum\limits_{n_1\,n_2} p_{n_1} p_{n_2} \cos ^4 \left (\frac{\Omega_{n_1,\,n_2} t}{2}\right ),\\
  C=\sum\limits_{n_1\,n_2} p_{n_1} p_{n_2} \sin ^4 \left (\frac{\Omega_{n_1,\,n_2} t}{2}\right )\>,\\
 \nonumber D= \sum\limits_{n_1\,n_2} p_{n_1} p_{n_2} (n_1+1) (n_2+1) S^2_{n_1,\,n_2}\>,\\
 \nonumber E=-\frac{1}{4} \sum\limits_{n_1\,n_2} p_{n_1} p_{n_2} \sin ^2 \left (\Omega_{n_1,\,n_2} t\right ),
\end{eqnarray}
where
$$ O_{n_1,\,n_2} = \langle n_1 \mid \langle n_2 \mid O \mid n_1 \rangle \mid n_2 \rangle.$$

The results of numerical calculations of entanglement on the basis of the formulae (5) and (6) are shown in Fig.1. In this the
 dependencies of the Peres-Horodecki parameter $ \varepsilon$ versus dimensionless time $gt$ is presented for models with
 $\overline{n}_1 = \overline{n}_2 = 0.3$ (dashed line) and $\overline{n}_1 = \overline{n}_2 = 1$ (solid line). Comparing
 curves with  that for one-photon \cite{Kim}  and  degenerate two-photon \cite{Zhou2} interaction, we find that contrary to
  above mentioned models
   the entanglement for considered model takes place  for all instances  only for sufficiently large  intensities of the
 input cavity modes. The  entanglement through  nondegenerate two-photon interaction reveals the nonlinear behaviour
 in a similar way to
 the degenerate interaction  but its value is less than that for aforementioned model with same values of input intensity.

2. If both atoms are initially in the ground state, i.e. the atomic state is  $\mid -, - \rangle $, then,the matrix elements
 of the atomic density operator (4) are
\begin{eqnarray}
\nonumber A= \sum\limits_{n_1\,n_2} p_{n_1} p_{n_2} n_1(n_1-1) n_2(n_2-1) C^2_{n_1-1,\,n_2-1}\>, \\
 B=C=E=\sum\limits_{n_1\,n_2} p_{n_1} p_{n_2} n_1 n_2 S^2_{n_1-1,\,n_2-1}\>,\\
 \nonumber  D=-\frac{1}{4} \sum\limits_{n_1\,n_2} p_{n_1} p_{n_2} [n_1 n_2 C_{n_1-1,\,n_2-1} + 1]^2.
 \end{eqnarray}
 The time evolution of the entanglement in this case is shown in Fig.2  for the same values of input intensities of the cavity modes.
 We observe again the difference between considered and degenerate two-photon process. If two atoms
 initially both in ground state, they only become slightly
 entangled through degenerate two-photon processes \cite{Zhou2} but for nondegenerate interaction the value of
 Peres-Horodecki parameter $ \varepsilon$ compares favourably with that for previous case when atoms are initially in the state
 $\mid+, - \rangle$. With same input intensities for considered model
 the maximum values of $\varepsilon$ are much larger than those for degenerate two-photon model.

3. For both atoms in the initial excited state $\mid +, + \rangle $ we have
\begin{eqnarray}
\nonumber A= \sum\limits_{n_1\,n_2} p_{n_1} p_{n_2} [(n_1+1)(n_2+1) C^2_{n_1+1,\,n_2+1} +1]^2\>, \\
 B=C=E=\sum\limits_{n_1\,n_2} p_{n_1} p_{n_2} (n_1+1)(n_2+1) S^2_{n_1+1,\,n_2+1}\>,\\
 \nonumber  D=\sum\limits_{n_1\,n_2} p_{n_1} p_{n_2} (n_1+1)(n_1+2)(n_2+1)(n_2+2) C_{n_1+1,\,n_2+1}^2.
 \end{eqnarray}
 Numerical calculations with using Eqs.(3),(8) reveals that there is no entanglement in the considered case in accordance with
 that for one-photon and degenerate two-photon processes.

 4. Let consider the atom-atom  entanglement when the initial atomic state is mixed so that the initial density operator are
 \begin{eqnarray}
\rho_A(0) = \prod\limits_{i=1,2} [\lambda \mid + \rangle_i \langle +\mid +(1-\lambda) \mid-\rangle_i\langle-\mid,
 \end{eqnarray}
where $\lambda/(1-\lambda) =\exp(\omega_0/k_B T).$
  M.Kim et al.\cite{Kim} and L.Zhou et al. \cite{Zhou2} have shown that, with  even a very small amount of mixture $\lambda$,
  atomic entanglement is washed  out  both for one-photon and degenerate two-photon interaction. Combining the results
  for three previous cases one can calculate the entanglement for initial mixed atomic state. The time dependence of
 $\varepsilon$ for model with
 $\lambda=0.01$ (solid line) and $\lambda=0.05$ (dashed line) has presented in Fig.3 for  $\overline{n}_1 =
 \overline{n}_2 = 1$.
 Note, that $\lambda=0$ corresponds to the atoms in the ground state. We find that with $\lambda$ increasing further to 0.09 ,
 the entanglement is vanished.

 So, we have shown that two identical atoms can be entangled through nonlinear nondegenerate
 two-photon interaction with two-mode thermal field with the exception of the case when both atoms are excited. For certain initial pure atomic
  states the two-atom entanglement
 may be stronger
 than that produced in one-photon and degenerate two-photon processes. For atoms are initially prepared in a
 thermal mixture of the ground and excited states the  entanglement is greatly decreases with increasing  the weight
 of the excited state in the mixture.

 The author is indebted to Prof. V.L.Derbor for discussions. This work was supported by RFBR grant 04-02-16932.
 
 \newpage

\begin{figure}
\centering\includegraphics[width=3.0in]{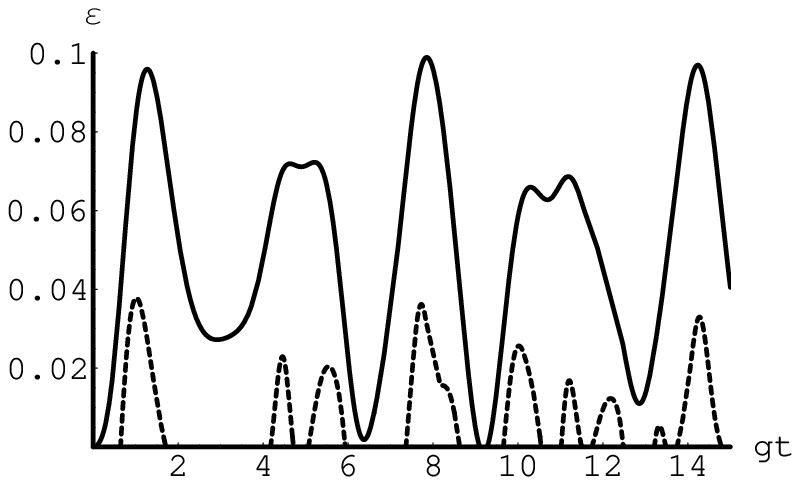} \caption{Atom-atom entanglement induced by interaction with two-mode thermal field when the atoms are
 initially prepared in $ \mid +, -\rangle $ for $\overline{n}_1 = \overline{n}_2 = 0.3$ (dashed line) and
  $\overline{n}_1 = \overline{n}_2 = 1$ (solid line). }
\end{figure}
\begin{figure}
\centering\includegraphics[width=3.0in]{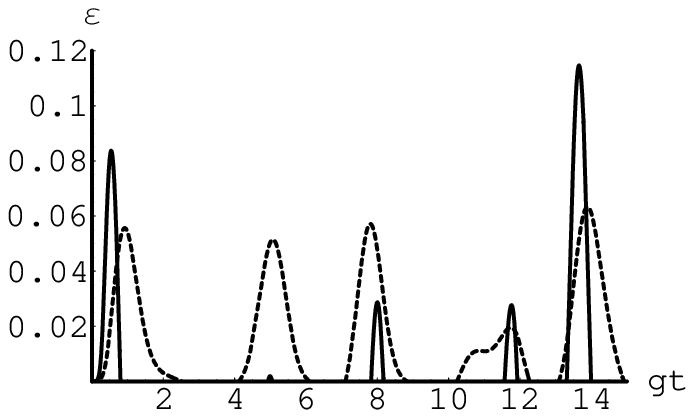} \caption{Atom-atom entanglement induced by interaction with two-mode thermal field when the atoms are
 initially prepared in $ \mid -, -\rangle $ for $\overline{n}_1 = \overline{n}_2 = 0.3$ (dashed line) and
  $\overline{n}_1 = \overline{n}_2 = 1$ (solid line). }
\end{figure}
\begin{figure}
\centering\includegraphics[width=3.0in]{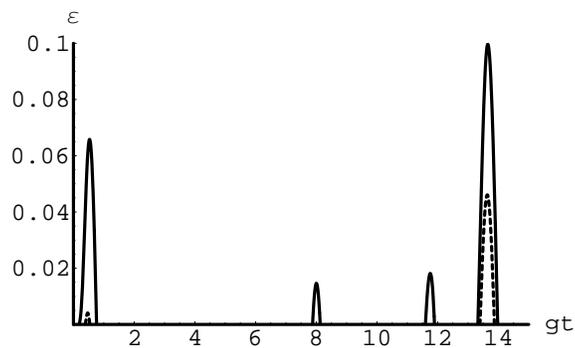} \caption{Atom-atom entanglement induced by interaction with two-mode thermal field when the atoms are
 initially prepared in mixed state with  $\lambda=0.01$ (solid line) and $\lambda=0.05$ (dashed line)
 for   $\overline{n}_1 = \overline{n}_2 = 1$ . }
\end{figure}







\end{document}